\documentclass[aps,prl,twocolumn,showpacs,groupedaddress]{revtex4}
\usepackage{graphicx}  
\usepackage{dcolumn}   
\usepackage{bm}        
\usepackage{amssymb}   
\begin{document}

\def\ra{{\rightarrow}}
\def\a{{\alpha}}
\def\b{{\beta}}
\def\l{{\lambda}}
\def\eps{{\epsilon}}
\def\T{{\Theta}}
\def\t{{\theta}}
\def\co{{\cal O}}
\def\car{{\cal R}}
\def\caf{{\cal F}}
\def\cs{{\Theta_S}}
\def\pr{{\partial}}
\def\tri{{\triangle}}
\def\na{{\nabla }}
\def\S{{\Sigma}}
\def\s{{\sigma}}
\def\sp{\vspace{.05in}}
\def\hs{\hspace{.25in}}
\def\rs{\vspace{-.08in}}

\newcommand{\be}{\begin{equation}} \newcommand{\ee}{\end{equation}}
\newcommand{\bea}{\begin{eqnarray}}\newcommand{\eea}
{\end{eqnarray}}


\begin{titlepage}

\title{\Large\bf Towards Nonperturbation Theory of Emergent Gravity}

\author{Supriya Kar}
\affiliation{Department of Physics \& Astrophysics, University of Delhi, New Delhi 110 007, India}

\begin{abstract}
We investigate an emergent gravity in $(4+1)$ dimensions underlying a geometric torsion ${\cal H}_3$ in $1.5$ order formulation. We show that an emergent pair-symmetric $4$th order curvature tensor, underlying a NS field theory, governs a torsion free geometry and may be identified with the Riemann type tensor. Interestingly a pair anti-symmetric $4$th order is shown to incorporate a dynamical correction underlying a ${\cal H}_3$ propagation and is identified as a non-perturbation (NP) correction. In particular the NP-dynamics is governed by the $U(1)$ gauge invariant ${\cal F}_4$ in second order underlying onshell NS field in first order. It is shown that a NP-theory of quantum gravity is elegantly described with an axion, and hence a quintessence, coupling to the Riemannian geometries. The curvatures are appropriately worked out to argue for $d$$=$$12$ emergent $F$-theory. Investigation reveals that a pair of $(M{\bar M})_{10}$-brane is created across an event horizon in $F$-theory. We show that an emergent $M$-theory in a decoupling limit may be identified with the bosonic sector of $N=1$ supergravity in $d$$=$$11$.
\end{abstract}

\pacs{11.25.-w, 11.25.yb, 11.25.Uv, 04.60.cf} 
\maketitle
\end{titlepage}

\noindent
{\it\bf Introduction:} Black holes are described in Einstein gravity which is governed by a metric dynamics on a Riemannian manifold. Interestingly the stringy versions of these macroscopic black holes have been obtained in a low energy limit of a string effective action  \cite{candelasHS,freed,gibbons-maeda-NPB,garfinkle,giddings-strominger}.

\sp
\noindent
In the past there are attempts to use a gauge principle for the Riemannian geometry \cite{wilczek}. The quantum effects to Einstein gravity can also be addressed non-perturbatively using the strong-weak coupling duality in ten dimensional superstring theories \cite{ashoke-sen-ijmpa}. A non-perturbative quantum effect is believed to be sourced by the compactified extra space dimension(s) to the stringy vacua. In particular the type IIA superstring theory in a strong coupling limit is known to incorporate an extra spatial dimension on $S^1$ and has been identified with $d$$=$$11$ non-perturbation $M$-theory. Generically $M$-theory has been shown to be identified with the stringy vacua in various dimensions \cite{witten-NPB}. $M$-theory in a low energy limit is known to reduce to the $N$$=$$1$ supergravity (SUGRA) in eleven dimensions \cite{cremmer-julia-scherk}. Matrix theory formulations \cite{BFSS-PRD,IKKT-NPB} have enhanced knowledge on the non-perturbation $M$-theory. However a complete $M$-theory and its intrinsic power to explore new physics has not been fully explored \cite{schwarz-PLB}.

\sp
\noindent
In the article we attempt to obtain an action for the $M$-theory in $d$$=$$11$. It is argued that the dynamics of a geometric torsion ${\cal H}_3$ in $1.5$ order can be a potential candidate to describe the $M$-theory. Investigation reveals that NP-theory of ${\cal H}_3$ is an emergent phenomenon on a fat brane for onshell NS field in first order \cite{abhishek-JHEP,abhishek-PRD,abhishek-NPB-P}. 

\sp
\noindent
Arguably an emergent NP-theory necessitates a hint for a {\it fundamental}, or $4$-{\it form}, theory in $d$$=$$12$ which may correspond to the $F$-theory \cite{vafa}. It is shown that a quintessence axionic scalar incorporates a NP-correction hidden to the torsion free semi-classical geometries. Thus a $D$-instanton, known as a fundamental building block in $F$ theory, may be a potential candidate to describe the quintessence and hence the dark energy in universe.

\sp
\noindent
{\it\bf Pair production scenarios:} We begin by briefly revisiting the Schwinger pair production mechanism \cite{schwinger}. The novel idea elegantly describes a $(e^+e^-)$ pair production at a quantum field theory (QFT) vacuum by a photon. The mechanism is an emergent NP-tool and is believed to be instrumental to describe diverse quantum phenomenon in gravitation and cosmology. In fact the tool was explored to explain Hawking radiation phenomenon \cite{hawking}, where an incident photon generates a pair of charged particle/anti-particle at the event horizon of a black hole. Furthermore the idea was applied to the open strings pair production \cite{bachas-porrati} and to investigate the production of a pair of $(D{\bar D})_9$ at the cosmological horizon \cite{majumdar-davis}. An extra spatial dimension on $S^1$ has been argued to unfold in $d$$=$$10$ type IIA superstring theory in a strong coupling and is believed to describe the $M$-theory in $d$$=$$11$.

\sp
\noindent
In the context an emergent NP-theory on a stringy pair by a Kalb-Ramond (KR) form quanta on a $D_4$-brane has been formulated in a collaboration \cite{abhishek-JHEP,abhishek-PRD,abhishek-NPB-P}. Importantly a generic vacuum geometry in GTR, $i.e.$ a Kerr black hole, has been obtained in the NP-decoupling limit from a degenerate Kerr family \cite{sunita-NPB}. The NP-formulation has further been explored to obtain the semi-classical Kerr-Newman black hole at the expense of the degeneracy in the quantum theory \cite{sunita-IJMPA}.

\sp
\noindent
{\it\bf Emergent gravity in ${\mathbf 5}$D:}
Interestingly the emergent gravity patches, underlying a non-linear higher form field quanta, were (spin) projected using a discrete matrix to describe an appropriate causal geometry in a low energy (GTR) limit \cite{priyabrat-EPJC,priyabrat-IJMPA}. The emergent gravity patches have formally been viewed as a lower dimensional $D_p$-brane correction to a low energy string vacuum. Since a $D_p$-brane carries Ramond-Ramond (RR) charge, the correction turns out to be non-perturbative \cite{polchinski}. 

\sp
\noindent
It was argued that an emergent pair breaks the supersymmetry due to the exchange 
of closed string modes between the vacuum created gravitational pair of $(3{\bar 3})$-brane across the cosmological horizon \cite{abhishek-JHEP,richa-IJMPD}. Thus a gravitational pair is described on a fat brane underlying a massive NS field dynamics in a first order theory. In fact, a quintessence scalar field acts as an extra (fifth) hidden dimension between a pair. It determines the thickness of a fat brane. Alternately a quintessence local degree, underlying the closed string exchange between the gravitational pair, is absorbed by a $D_3$-brane to make it a fat $3$-brane. In the context a vacuum refers to a background black hole defined with an open string metric ${\tilde{G}}_{\mu\nu}=\big ( g_{\mu\nu}-B_{\mu}^{(NS)\lambda}B^{(NS)}_{\lambda\nu}\big )$ on a $D_4$-brane, where $g_{\mu\nu}$ is a constant metric. Thus the open string metric is sourced by a constant, or a global mode of, NS field \cite{seiberg-witten} in the string bulk. In addition the Poincare duality in the five dimensional world-volume gauge theory ensures a two form (KR) dynamics on a $D_4$-brane. Thus the non-linear $U(1)$ gauge dynamics of a $D_4$-brane in presence of a background metric may be envisaged with a constant NS field and a local KR field. It is described \cite{abhishek-JHEP} by
\rs
\be
S={{-1}\over{(8\pi^3g_s){\alpha'}^{3/2}}}\int d^5x {\sqrt{-{\tilde G}^{({\rm NS})}}}\ H_{\mu\nu\lambda}H^{\mu\nu\lambda}\ ,\label{p-1}
\ee

\vspace{-.04in}
\rs
\noindent
where ${\tilde{G}}=$$\det{\tilde{G}}_{\mu\nu}\ {\rm and}\ H_{\mu\nu\lambda}$$=$$\big ( \nabla_{\mu}B^{(KR)}_{\nu\lambda}$$+{\rm cyclic}\big )$.

\sp
\noindent
Generically the idea is to explore a higher form $U(1)$ gauge theory, and hence its geometric effects, on an appropriate $D_p$-brane. It fact a higher form gauge theory is non-linear and becomes sensible in higher dimensions. It may provide a clue to unfold a NP-theory of gravity in higher dimensions such as $d$$=$$11$ $M$-theory \cite{witten-NPB,schwarz-PLB}. A priori, for a simplicity, we begin with a two form gauge theory on a $D_4$-brane in $d$$=$$10$ type IIA superstring theory. Nevertheless, a five dimensional construction is a minimal extension to the GTR. It is believed that GTR is fundamental and is an emergent NP-phenomenon. As a bonus, the NP-formulation takes into account the quintessential cosmology. The NP-tool has been explored in the recent past to explain the observed accelerated rate of expansion of universe \cite{priyabrat-EPJC} and the origin of dark energy \cite{priyabrat-IJMPA}.

\sp
\noindent
{\it\bf NS field dynamics:}
An emergent fat brane evolves with a dynamical NS field which is obtained at the expense of the KR field dynamics on a $D_4$-brane. Thus a covariantly constant NS field, $i.e.\ \nabla_{\alpha}B_{\mu\nu}^{(NS)}=0$, on a $D_4$-brane becomes dynamical on a fat brane. The $U(1)$ gauge invariant ${\cal H}_3$ in an emergent NS theory is interpreted as a geometric torsion ${\cal H}_3$ and is given by
\rs
\bea
&&{\cal H}_{\mu\nu\lambda}={\cal D}_{\mu}B^{(NS)}_{\nu\lambda}+{\cal D}_{\nu}B^{(NS)}_{\lambda\mu}+{\cal D}_{\lambda}B^{(NS)}_{\mu\nu}
\ ,\qquad\; {}\nonumber\\
{\rm where}&&{\cal D}_{\lambda}B^{(NS)}_{\mu\nu}={1\over2}\left ({H_{\lambda\mu}{}}^{\rho}B^{(NS)}_{\nu\rho}+ {H_{\lambda\nu}{}}^{\rho}B^{(NS)}_{\rho\mu}\right ).\label{gtorsion-1}
\eea

\rs
\noindent
A constant NS field background in a closed string becomes a nontrivial geometric torsion at the expense of the the KR field dynamics, $i.e.\ {\cal D}_{\lambda}B_{\mu\nu}^{(KR)}=0$. It implies that the modified ${\cal D}_{\mu}$ is uniquely fixed. In fact the absorption of KR field quanta, in a background black hole, has given birth to an emergent quantum theory of gravitation on a stringy pair of $(3{\bar 3})$-brane across an event horizon \cite{abhishek-PRD}. Interestingly a plausible NP-theory of gravity underlying an emergent scenario enforces an iterative correction: 
$H_3\rightarrow {\cal H}_3$. It leads to an exact description with a $B^{(NS)}_2$ perturbation:
\rs
\be
{\cal H}_{\mu\nu\lambda}=H_{\mu\nu\rho}B_{\lambda}^{(NS)\rho}+ H_{\mu\nu\alpha}B_{\rho}^{(NS)\alpha} B_{\lambda}^{(NS)\rho}+\dots\label{gtorsion-2}
\ee

\rs
\noindent
For an infinitesimal perturbation: ${\cal H}_3\rightarrow H_3$. Thus an exact perturbation NS field theory may equivalently be described by a NP-theory of gravity for onshell NS field.

\sp
\noindent
Generically there are three distinct dynamical aspects of the same description in $d$$=$$5$ and they are: (i) a KR perturbation $U(1)$ theory defined with a covariant derivative $\nabla_{\mu}$ on a $D_4$-brane, (ii) a NS perturbation $U(1)$ theory defined with a modified covariant derivative ${\cal D}_{\mu}$ on an emergent fat brane and describes semi-classical geometries, (iii) a geometric torison NP-theory of emergent quantum gravity on a vacuum created pair of $(3{\bar 3})$-brane underlying an enhanced gauge group $U(1)\times U(1)_{\rm NP}$. An additional $U(1)_{\rm NP}$ in a NP-theory is only sensible in a second order while the KR and NS perturbation theories are first order. The original and the emergent perturbations are respectively described by the $U(1)$ gauge invariant field strengths $H_3$ in eq(\ref{p-1}) and ${\cal H}_3$ in eq(\ref{gtorsion-1}). However the NP-theory breaks the gauge invariance perturbatively which is evident from the field strength expression in eq(\ref{gtorsion-2}). The perturbative gauge invariance has been realized \cite{abhishek-JHEP,priyabrat-EPJC} in presence of a symmetric fluctuation: $f_{\mu\nu}={\bar{\cal H}}_{\mu\alpha\beta}{{\cal H}^{\alpha\beta}{}}_{\nu}$, where ${\bar{\cal H}}_3=(2\pi\alpha'){\cal H}_3$. Thus a dynamical NS field modifies the constant background metric $G_{\mu\nu}^{(NS)}$ into a dynamical metric on an emergent fat brane. The modified metric is given by
\rs
\be
G_{\mu\nu}=\Big ( g_{\mu\nu}-B_{\mu}^{(NS)\lambda}B^{(NS)}_{\lambda\nu} + {\bar{\cal H}}_{\mu\lambda\rho}{{\cal H}^{\lambda\rho}}_{\nu}\Big )\ . \label{gauge-metric}
\ee

\rs
\noindent
There are two interesting observations in an emergent scenario. 
Firstly the metric $G_{\mu\nu}$ dynamics is governed by an underlying NS field theory. It generates a geometric torsion and hence 
incorporates an intrinsic angular momentum in an emergent vacuum. As a result an emergent gravity naturally governs the Kerr black hole as a vacuum geometry \cite{sunita-NPB,sunita-IJMPA}. Secondly, the metric dynamics can not be perceived directly in a first order theory and urges a need for a second order formulation. Importantly the gauge invariance shall be shown to be restored in a second order for onshell NS field in the paper. As a result, the ${\cal H}_3$ is perceived as a gauge potential underlying a NP-dynamical correction to the perturbative NS field theory in first order and hence the complete NP-theory is governed in $1.5$ order by ${\cal H}_3$.

\sp
\noindent
At a first sight it may imply that the modified metric dynamics is primarily governed by a propagating ${\cal H}_3$. However a geometric torsion theory, being a quantum description, cannot be identified with a metric tensor theory. This is due to a fact that the (pseudo) Riemannian geometry is sourced by a metric tensor (GTR). The apparent puzzle is resolved with an emergent metric which is a semi-classical phenomenon as it is realized in a first order perturbation theory of NS field. The complete NP-theory in $1.5$ order is an emergent quantum gravity description underlying a geometric torsion for the onshell NS field. Thus the notion of a dynamical metric is not sensible due to the prevailing geometric torsion dynamics in a NP-theory of quantum gravity in $d$$=$$5$. A propagating torsion is known to break the space-time continuum and hence there are no closed paths. 

\sp
\noindent
In the paper we explicitly work out a NP-dynamical correction in $d$$=$$5$ emergent scenario which in turn is a key to address an emergent $M$-theory derived from a {\it fundamental}, presumably $F$-theory, in $d$$=$$12$. It has been shown that the GTR emerges as NP-phenomenon on a stable gravitational pair of $((3{\bar 3})$-brane \cite{abhishek-JHEP,abhishek-PRD,abhishek-NPB-P}. A fact that GTR is a fundamental unit, or a $3$-brane in the formulation, allows one to believe in $d$$=$$12$ {\it fundamental} theory which is a primary theme in this article. It is shown that the dynamical-NP couples to the Riemann tensor. The coupling is realized in terms of the Riemann left and right duals. Possibly it signals the intrinsic role played by a geometric torsion in the GTR.

\sp
\noindent
{\it\bf Emergent curvature tensors} {$\big ({{{\cal H}_{\mu\nu}{}}^{\lambda}},\ {{\cal K}_{\mu\nu\lambda\rho}},\ {{\cal L}_{\mu\nu\lambda\rho}}\big )$:}
The commutator of a modified covariant derivative is worked out to obtain the generic curvature tensors \cite{abhishek-JHEP}. It was shown that 
an emergent curvature is described on a vacuum created gravitational pair of $(3{\bar 3})$-brane \cite{abhishek-PRD}. The commutator operations yield:
\rs
\bea
&&\left [ {\cal D}_{\mu}\ ,\ {\cal D}_{\nu}\right ]\psi\ =\  -2\ {{\cal H}_{\mu\nu}{}}^{\lambda}\ {\cal D}_{\lambda}\psi\nonumber\\
{\rm and}&&\left [ {\cal D}_{\mu}\ ,\ {\cal D}_{\nu}\right ]A_{\lambda}\ =\  \left ({{\cal K}_{\mu\nu\lambda}{}}^{\rho}
\ +\ {{\cal L}_{\mu\nu\lambda}}^{\rho}\right )A_{\rho}\ ,\label{new-curvature}
\eea

\rs
\noindent
where ${\cal H}_3$ ensures a NS field dynamics in an emergent first order perturbation theory. 
It provokes thought to believe that the quanta of NS field in superstring theory can govern an emergent graviton. 

\sp
\noindent
Furthermore the coupling of ${\cal H}_3$ to the {\it dynamics} ${\cal D}_{\lambda}\psi$ rather than the field $\psi$ in eq(\ref{new-curvature}) is a new phenomenon. A non-zero ${\cal H}_3$ signifies a difference in energy between (emergent) quantum gravity and classical (Einstein) gravity. In fact ${\cal D}_{\mu}\psi\neq 0$, $i.e.$ the dynamical scalar, is known to be pivotal to incorporate a quantum NP-correction to the GTR \cite{priyabrat-EPJC,priyabrat-IJMPA}. Thus a dynamical correction to the GTR plays a significant role in quantum cosmology and is indeed a powerful tool to explain the accelerated rate of expansion of universe. The inherent (quantum) dynamics of the scalar field makes the emergent gravity a natural candidate to describe the quintessence which is believed to govern the dark energy.

\sp
\noindent
The 4th order emergent curvature tensors (\ref{new-curvature}) are reducible and they play a significant role in a NP-theory. They are believed to possess a key to a complete NP-theory such as $M$ theory. Generically ${\cal K}_{\mu\nu\lambda\rho}$ can be re-expressed as a pair symmetric $(S)$ and  pair non-symmetric $({\tilde A})$ under an interchange of first and second pair of indices, $i.e.\ {\cal K}^{(S)}_{\mu\nu\lambda\rho}+{\cal K}^{({\tilde A})}_{\mu\nu\lambda\rho}$. Explicitly all three emergent curvatures are:
\rs
\bea
{{\cal L}_{\mu\nu\lambda}}^{\rho}&=&\Gamma^{\rho}_{\mu\sigma} {{\cal H}^{\sigma}}_{\nu\lambda}+\Gamma^{\sigma}_{\nu\lambda} {{\cal H}^{\rho}}_{\mu\sigma} - \Gamma^{\sigma}_{\mu\lambda} {{\cal H}^{\rho}}_{\nu\sigma} -\Gamma^{\rho}_{\nu\sigma} {{\cal H}^{\sigma}}_{\mu\lambda}\nonumber\\
{\cal K}^{(S)}_{\mu\nu\lambda\rho}&=&{{\cal H}_{\mu\lambda}}^{\sigma}{\cal H}_{\nu\sigma\rho}-{{\cal H}_{\nu\lambda}}^{\sigma}
{\cal H}_{\mu\sigma\rho}\ ,\nonumber\\
{\cal K}^{({\tilde A})}_{\mu\nu\lambda\rho}&=&\partial_{\mu}{\cal H}_{\nu\lambda\rho} -\partial_{\nu} {\cal H}_{\mu\lambda\rho}\ ,\label{g-curvature}
\eea

\rs
\noindent
where  $\Gamma^{\rho}_{\mu\sigma}$ denote the Christoeffel connections sourced by the background metric ${\tilde G}_{\mu\nu}$. Thus ${{\cal L}_{\mu\nu\lambda}}^{\rho}$ describes a coupling between the GTR and an emergent gravity. It ensures a non-propagating geometric torsion and hence a torsion free geometry. An irreducible tensor is worked out to yield:
\rs
\be
{\cal L}_{\mu\lambda}=\Gamma^{\nu}_{\mu\rho} {{\cal H}^{\rho}}_{\nu\lambda}+\Gamma^{\rho}_{\nu\lambda} {{\cal H}^{\nu}}_{\mu\rho}- \Gamma^{\nu}_{\nu\rho} {{\cal H}^{\rho}}_{\mu\lambda}=-{\cal L}_{\lambda\mu}\ .\label{gauge-65}
\ee

\rs
\noindent
The Lorentz scalar $\big ({\cal L}_{\lambda\mu}{\cal L}^{\lambda\mu}\big )$ involves quadratic power in geometric torsion and Christoeffel connection(s). Since the  perturbation parameter is infinitesimally small, quadratic and higher powers become insignificant. A priori, it may imply that an emergent gravity in an all order perturbation theory is defined in a decoupling limit of two distinct connections. However a NP-theory incorporates the coupling of connections defined with the Lorentz scalar 
$\big ({\cal L}_{\lambda\mu}{\cal L}^{\lambda\mu}\big )$.

\sp
\noindent
The emergent curvature tensor ${\cal K}^{(S)}_{\mu\nu\lambda\rho}$ of order four (\ref{g-curvature}) shares all the properties of a Riemann tensor under the interchange of indices within a pair and with pairs. In addition the pair-symmetric curvature is sourced by a NS field propagation in an emergent perturbation theory in first order. As explained, it does not sense the propagation of a ${\cal H}_3$ which is a second order phenomenon. It describes a torsion free geometry. Thus a pair-symmetric tensor may be identified with the Riemann type tensor ${\cal R}_{\mu\nu\lambda\rho}$.

\sp
\noindent
{\it\bf Non-perturbative correction:}
The ${\cal H}_3$ dynamics may be re-expressed with a pair anti-symmetric $(A)$ tensor:
\rs
\be
{\cal K}^{({\tilde A})}_{\mu\nu\lambda\rho}\ =\ {1\over2}\Big ({1\over{\sqrt{2\pi\alpha'}}}\ {\cal F}_{\mu\nu\lambda\rho} \ +\ {\cal K}^{(A)}_{\mu\nu\lambda\rho}\Big )\ ,\label{gauge-7777}
\ee

\rs
\noindent
where ${\cal F}_{\mu\nu\lambda\rho}=4{\sqrt{2\pi\alpha'}}\ {\cal D}_{[\mu}{\cal H}_{\nu\lambda\rho]}$.
The second term is described by an irreducible curvature: 
\rs
\be
{\cal K}^{(A)}_{\mu\lambda}=\ 2\big ({\cal L}_{\mu\lambda} -{\cal D}_{\rho}{{\cal H}^{\rho}{}}_{\mu\lambda}\big )\ 
\longrightarrow 2\ {\cal L}_{\mu\lambda} 
\ .\label{gauge-77}
\ee

\rs
\noindent
Thus onshell NS form implies that the two form ${\cal K}^{(A)}_{\mu\nu}$ becomes trivial in the action. Then the Lorentz scalar:
\rs
\be
{\cal K}^{({\tilde A})}_{\mu\nu\lambda\rho} {\cal K}_{(A)}^{\mu\nu\lambda\rho}\ =\ {1\over{8\pi\alpha'}} {\cal F}_{\mu\nu\lambda\rho}{\cal F}^{\mu\nu\lambda\rho}
\ .\label{gauge-792}
\ee

\rs
\noindent 
Irreducibility of a pair-symmetric tensor (\ref{g-curvature}) is worked out to obtain: ${\cal K}_{\mu\nu}=-\big ({\cal H}_{\mu\rho\lambda}{{\cal H}^{\rho\lambda}{}}_{\nu}\big )$ and ${\cal K}=-{\cal H}_3^2$. They describe torsion free geometries and may 
formally be identified with the Ricci type tensors: ${\cal R}_{\mu\nu}$ and ${\cal R}$. They lead to an emergent metric in a first order perturbation gauge theory which is precisely governed by a NS field dynamics on a fat brane. However a dynamical correction (\ref{gauge-77}) in second order ensures that a NP-theory is completely perceived in $1.5$ order formulation

\sp
\noindent
{\it\bf Topological couplings:}
A priori there are two topological couplings between a constant NS form and a dynamical KR form in the world-volume gauge theory on a $D_4$-brane. Both of them turn out to be a total divergence in the $d$$=$$5$ bulk (B) though they contribute significantly to the boundary (BD) theory. They are given by
\rs
\bea
{1\over{{\kappa'}^4}}\int_{\rm B} H_3\wedge\Big (B_2^{(NS)}+{\bar F}_2\Big )={1\over{\kappa^4}}\int_{\rm BD} B_2^{(KR)}\wedge {\bar{\cal F}}_2, \label{topological-1}
\eea

\rs
\noindent
where $\kappa'$ and $\kappa$ are respectable couplings in $d$$=$$5$ and $d$$=$$4$. They possess a dimension of length.
Remarkably the boundary action identifies with the $BF$-topological theory. For instance see ref.\cite{BF-1}. In an emergent scenario the topological 
coupling is re-expressed as:
\rs
\be
{1\over{\kappa'}^4}\int_{\rm B} B_2^{(KR)}\wedge {\cal H}_3\ =\ {1\over{\kappa}^4}\int_{\rm BD} B_2^{(KR)}\wedge B_2^{(NS)}\ .\label{topological-2}
\ee

\rs
\noindent
In addition the NS and RR couplings are known to incorporate a notion of branes within a brane. The novel idea has been pivotal to the formulation of two independent $d$$=$$10$ Matrix theories. They are: (i) $D$-particle as a building block for type IIA superstring \cite{BFSS-PRD} and 
(ii) $D$-instanton for type IIB superstring \cite{IKKT-NPB}.

\sp
\noindent
Though a topological coupling is trivial in the bulk, it regains significance in the boundary and is given by 
\rs
\be
{1\over{\kappa'}^3}\int_{\rm B}{\cal F}_4\wedge {\cal F}_1={1\over{\kappa}^2}\int_{\rm BD} {\cal H}_3\wedge d\psi
={{\Lambda}\over{\kappa}^3}\int_{\rm BD} \psi
\ ,\label{topological-3}
\ee

\rs
\noindent
where ${\cal H}_3$ turns out to be a $U(1)$ gauge potential and $\psi$ denotes an axion coupling. Non-zero $\Lambda=\big ({\cal E}^{\mu\nu\lambda\rho}\ {\cal F}_{\mu\nu\lambda\rho}\big )$ may be identified with a cosmological constant in the boundary theory. Its toplogical coupling to an axionic scalar field in $d$$=$$4$ theory is remarkable. It may imply that a topological coupling of an axion in $d$$=$$4$ plays an important role to describe a 
hidden quintessence.

\sp
\noindent
{\it\bf Gravity duals:}
The pair-symmetric $(S)$ and the pair non-symmetric  $({\tilde A})$ emergent curvatures 
tensors  of order four (\ref{g-curvature}) are worked out for their irreducibility to yield a curvature scalar ${\cal K}$$=$${\cal R}$ and ${\cal F}_4$ respectively. The four form has been shown to incorporate a NP-dynamical correction obtained via eqs(\ref{gauge-7777})-(\ref{gauge-792}). The form of ${\cal F}_4$ further ensures a propagating ${\cal H}_3$. Thus an effective action for a NP-theory of emergent quantum gravity on a 
gravitational pair of $(3{\bar 3})$-brane is given by
\rs
\be
S_{EG}={1\over{{\kappa'}^3}}\int d^5x {\sqrt{-g}}\; \Big ( {\cal R}-{1\over{2\cdot 4!}}\ {\cal F}_4^2\Big )\ ,\label{gauge-70}
\ee

\rs
\noindent
where the $g_{\mu\nu}$ signature is $(-,+,+,+,+)$ and the string slope parameter $\alpha'$ is used to re-express ${\kappa'}^2=(8\pi\alpha')$. The first term governs a torsion free geometry and a second term incorporates a dynamical correction.
Remarkably the emergent gravity action formally resembles to the bosonic sector of $d$$=$$11$ SUGRA  \cite{cremmer-julia-scherk}. The emergent NP-theory is re-expressed as: 
\rs
\be
S_{EG}={1\over{{\kappa'}^3}}\int_B {\sqrt{-g}}\Big ( {\cal R} -{1\over{96}}\ {\cal F}_4 \wedge {}^{\star}d\psi\Big )
+{{\Lambda}\over{{\kappa}^3}}\int_{\rm BD} \psi\label{Egravity-dual}
\ee

\rs
\noindent
The equivalence, between (\ref{gauge-70}) and (\ref{Egravity-dual}), reconfirms a topological coupling and hence ${\cal F}_4$ in eq(\ref{gauge-7777}) modifies to: ${\sqrt{2\pi\alpha'}}\big ( d{\cal H}_3-{\cal H}_3\wedge d\psi\big )$. A dual action ensures a dynamical axionic scalar 
with metric signature $(+,+,+,+,-)$ and is given by
\rs
\be
S_{EG}^{Dual}\equiv\int_{\rm B}{\sqrt{-g}}\Big ( {\cal R}+{1\over2}\big ({\cal D}\psi\big )^2\Big )\ +\ \int_{\rm BD} {\cal H}_3\wedge d\psi\ .\label{gauge-702}
\ee

\rs
\noindent
An axionic coupling, underlying a $D$-instanton, has been investigated to incorporate the quintessence effect \cite{priyabrat-EPJC,priyabrat-IJMPA}. The left (L) and right (R) duals of the Riemann type tensor are worked out to yield:
\rs
\bea
{\cal R}^{(L)}_{\mu\nu\lambda\rho}&=&\sqrt{{\pi\alpha'}\over2} {\cal F}_{\mu\nu\alpha\beta}\ {{\cal R}^{\alpha\beta}{}}_{\lambda\rho}\nonumber\\
{\rm and}\qquad {\cal R}^{(R)}_{\mu\nu\lambda\rho}&=&\sqrt{{\pi\alpha'}\over2} 
{{\cal R}_{\mu\nu}{}}^{\alpha\beta}\ {\cal F}_{\alpha\beta\lambda\rho}\ .\label{gauge-704}
\eea

\rs
\noindent
Remarkably the geometric duals reveal an intrinsic coupling of ${\cal F}_4$ with the Riemann type tensors in a NP-theory. The coupling ensures the presence of an axionic scalar in $d$$=$$4$ and may hint for a desirable quintessence correction to the GTR. Quintessence is known to be a candidate to describe the dark energy universe.

\sp
\noindent
{\it\bf Emergent ${\mathbf F}$-theory:}
We begin with an emergent NP-theory (\ref{gauge-70}) governed by the ${\cal H}_3$ gauge potential in $1.5$ order. Furthermore a $H_3$$=$$dB_2^{(KR)}$ is sourced by a string charge and is best described on a $D_5$-brane due to a CFT \cite{richa-IJMPD,deobrat-Springer}. A $D_5$-brane is sourced by a 6-form in NS-NS and in RR sectors of type IIB superstring. It re-ensures a string/five-brane duality in $d$$=$$10$ superstring theory. Thus the significance of a $3$-brane in an emergent gravity may demand a $d$$=$$12$ {\it fundamental} theory. Interestingly $F$ theory has been viewed as a re-formulation of type IIB superstring theory with $D$-instanton as its fundamental unit \cite{vafa}. The $SL(2,Z)$ symmetry in type IIB supergravity has been exploited by the author to obtain $d$$=$$12$ theory \cite{kar-NPB}. The field theoretic construction in $d$$=$$12$ leading to $F$-theory has widely been studied \cite{pope-CQG,ueno-JHEP}. 

\sp
\noindent
In fact a dynamical ${\cal H}_3$ decouples to leave behind a torsion free geometry. Analysis leading to a stable gravitational $3$-brane as a fundamental unit in an emergent gravity provides a clue to $F$ theory in $d$$=$$12$. 
In addition an emergent gravity sourced by a KR field may formally be generalized to obtain a space filling emergent pair of $(8{\bar 8})$-brane underlying a $D_9$-brane in type IIB superstring. In principle a $9$-brane would be governed by $F_{11}$$=$$dB_{10}$ and hence the $10$-form theory requires a minimum of $d$$=$$12$. 
The hint signifies the role of a constant $F_5$$=$$dB_4$ in an emergent gravity (\ref{gauge-70}). The constant turns out to be dynamical in six and higher dimensions. Importantly a constant in $d$$=$$5$ describes a nontrivial topological coupling in addition to a total divergence $\left ({\cal F}_4\wedge {\cal F}_4\wedge {\cal F}_4\right )$ in $d$$=$$12$. We set $(2\pi\alpha'=1=\kappa')$ for all expressions starting herewith in the article. Then the {\it fundamental} action is a priori given by
\rs
\be
S=\int_{12}\Big ({\sqrt{-{\hat g}}}\Big ( {\hat{\cal R}} -{1\over{48}}{\cal F}_4^2 - {1\over{240}} F_5^2\Big ) + 
B_4\wedge {\cal F}_4\wedge {\cal F}_4\Big )\label{form-1}
\ee

\rs
\noindent
Interestingly a gravitational $3$-brane is governed by the $F$ theory in $d$$=$$12$. The Ricci type scalar ${\hat{\cal R}}$ governs a NS field dynamics and is believed to describe an emergent metric (\ref{gauge-metric}) in 
$F$-theory. The 4-form signifies a NP-correction underlying a ${\cal H}_3$ propagation. However the $F_5$$=$$dB_4$ may source 
a gravitational $3$-brane in $F$-theory. Generically the $F$ theory (\ref{form-1}) may be re-expressed as:
\rs
\be
S=\int_{12}\Big ({\sqrt{-{\hat g}}}\Big ( {\hat{\cal K}} -{1\over{48}}{\cal F}_4^2 - {1\over{240}} F_5^2\Big ) + 
B_4\wedge {\cal F}_4\wedge {\cal F}_4\Big )\label{form-2}
\ee

\rs
\noindent
The first two terms possess their origin in an emergent NP-theory of gravity underlying a geometric torsion ${\cal H}_3$. The field 
strength $F_5$$=$$dB_4$ possibly ensures an emergent gravitational $3$-brane.

\sp
\noindent
{\it\bf Emergent ${\mathbf M}$-theory:}
The total local degrees in $F$-theory (\ref{form-2}) turns out to be $l$$=$$375$. It includes the NP-local degrees $l_{\rm NP}$$=$$120$ corresponding to ${\cal F}_4$. The remaining $255$ local degrees are sourced by the NS field with $l_{\rm NS}$$=$$45$ and a $4$-form field with $l_{\rm 4F}$$=$$210$. The value of $l_{\rm NP}$ decreases down the dimension of space-time with $l_{\rm NP}$$=$$l_{\rm NS}$ in $d$$=$$7$ and the minimal NP-local degree $i.e.\ l_{\rm NP}$$=$$1$ is in $d$$=$$5$. 

\sp
\noindent
Generically the NP-local degrees in $(d+1)$-dimensions is equal to the local degrees of NP-theory underlying the NS-field and geometric torsion field in $d$-dimensions, $i.e.\ l_{\rm NP}^{d+1}$$=$$l_{\rm NS}^d$$+$$l_{\rm NP}^d$. Furthermore $l_{\rm NS}^d$$=$$l_{\rm metric}^{d-1}$$+$$l_{\rm HS}$, where the HS (Higher essence, or dimensional, scalar field) local $l_{\rm HS}$$=$$1$. The two level correspondences between a form theory in $(d+1)$-dimensions, via a NP-theory in $d$-dimensions, to a metric theory in $(d-1)$-dimensions is remarkable. It is a potential tool to describe an emergent NP theory on a gravitational pair of $(M{\bar M})$-brane. For instance, the $d$$=$$12$ (four) form theory can be mapped to $d$$=$$11$  non-perturbation $M$-theory. It is plausible to speculate that the NP-theory may further be viewed as a metric theory possibly corresponding to a space filling brane in $d$$=$$10$ superstring theories. Similarly a $4$-form theory in $d$$=$$6$ may be via $d$$=$$5$ (non-supersymmetric) NP-theory can be mapped to the GTR. The detailed analysis is beyond the scope of this paper.

\sp
\noindent
The $F$-theory action on $S^1$ is worked out for the massless fields to describe a total $l$$=$$375$ local degrees on an emergent gravitational pair of $(M{\bar M})$-brane. The irreducible curvatures ${\hat{\cal K}}$, ${\cal F}_4=d{\cal H}_3$ and $F_5=dB_4$ (\ref{form-2}) are respectively reduced to a pair of curvatures $({\tilde{\cal K}}, {\cal F}_2^2)$, $({\cal F}_4, {\cal K})$ and $(F_4$$=$$dB_3, F_5)$ on $(M{\bar M})$-brane. The dimensionally reduced action a priori be given by
\bea
&&S=\int_M {\sqrt{-{\tilde g}}}\ \Big ( {\tilde{\cal K}}\ -{1\over{48}}{\cal F}^a_4N_{ab}{\cal F}^b_4\Big )\nonumber\\
&&+\int_{\bar M} {\sqrt{-{\tilde g}}}\ \Big ({\cal K}\ -{1\over{4}} {\cal F}_2^2\ -{1\over{240}} F_5^2\ \Big )\nonumber\\
&&+\int_{(M{\bar M})}\Big [ {\cal H}_3\wedge \Big ({\cal F}_4\wedge {\cal F}_4\; + {\cal F}_4\wedge F_4\; + F_4\wedge F_4\Big )\nonumber\\
&&\qquad\qquad\qquad\quad\quad +\ F_2\wedge \Big ({\cal F}_4\ +{\cal F}_4\Big )\wedge F_5\Big ],\label{form-3}
\eea

\rs
\noindent
where $N_{ab}$ describes a $(2\times 2)$ diagonal matrix. The second term on a $M$-brane ensures two $4$-forms: ${\cal F}_4$ and $F_4$. The matrix element $N_{11}$ ensures a non-canonical potential coupling only to the ${\cal F}_4$ which is sourced by the ${\cal F}_4$ in $F$-theory. Since the geometric torsion theory in $d$$=$$12$ does not couple to $F_5$, the element $N_{22}$ is a constant. The second term contains a NP-curvature sourced 
by a propagating ${\cal H}_3$ in addition to a gauge theoretic $F_4$. 

\sp
\noindent
Interestingly the bulk $F$-theory on $R\times S^{11}$ leads to a boundary description for an emergent $M$-theory on $S^{11}$ underlying a pair of $(M{\bar M})$ across an event horizon. However invoking a generic bulk (NS-field)/boundary (metric tensor) correspondence between the $F$-theory and the $M$-theory, the curvatures ${\hat{\cal K}}$, ${\cal F}_4$ and ${\cal F}_5$ in eq(\ref{form-2}) are respectively re-expressed as: $(R, \phi)$, $({\cal F}_4=[d{\cal H}_3-{\cal H}_3\wedge d\phi], {\cal K})$ and $(F_4, F_5)$ on an emergent pair, where $R$ and $\phi$ respectively denote the Ricci scalar and a HS-field. Needless to mention that $45$ local degrees of NS field underlying ${\hat{\cal K}}$ in $d$$=$$12$ are described in $d$$=$$11$ by $44$ local degrees of an emergent metric and one local degree of a scalar. In fact the $\phi$ field is an extra $12$-th transverse dimension between a pair. Then the $d$$=$$12$ NS field dynamics in eq(\ref{form-2}), under the bulk/boundary correspondence, may be re-expressed in terms of Riemannian geometry in $d$$=$$11$. The complete NP-dynamics on a pair is given by
\rs
\bea
S&=&\int_M \Big [{\sqrt{-G}}\ \Big ( R_G\ -\ {1\over{48}}{\cal F}^a_4N_{ab}{\cal F}^b_4\Big )\nonumber\\
&&\qquad\qquad +\ \Big ({\cal H}_3\wedge{\cal F}_4\wedge {\cal F}_4\ +\ {\cal B}_3\wedge F_4\wedge F_4\ \Big )\Big ]\nonumber\\
&&+\int_{\bar M}\ {\sqrt{-G}}\ \Big ( {\cal K}\ - {1\over2}({\cal D}\phi)^2\ -{1\over{240}} F_5^2\Big )\ ,\label{form-4}
\eea

\rs
\noindent
where $G_{\mu\nu}$, in the invariant-volume, denotes a dynamical (emergent) metric obtained under a bulk/boundary correspondence on a pair. The NP-action (\ref{form-4}) may a priori be identified with the $M$-theory \cite{witten-NPB}. Thus the $M$-brane dynamics within a pair effectively takes into account a dynamical NP-correction and may formally be argued to govern the $M$-theory. Since a gravitational pair is created across an event horizon in a NP-formulation \cite{abhishek-JHEP,abhishek-PRD}, the $\phi$-field in $d$$=$$11$ acts as an extra hidden (transverse) space dimension.

\sp
\noindent
Interestingly a varying thickness of fat brane can be fixed by freezing HS-local degree, $i.e.\ \phi\rightarrow \phi_0$. It disconnects an emergent $M$-brane in a pair from ${\bar M}$-brane, where the radius $R$ of $S^{11}$ takes a fixed value $\phi_0$. The emergent curvatures and hence the causal effects on a gravitational $M$-brane universe are de-linked from that on ${\bar M}$-brane, as the $d$$=$$12$ coordinate system breaks down in the limit $\phi\rightarrow\phi_0$. Intuitively the higher dimensional limit 
may imply that the polar angle $\theta\rightarrow 0$ on $S^{11}$. The apparent angular deficit angle $\theta$ under a wick rotation may be viewed as light-like. Thus the disconnected emergent causal geometries are separated by a light-like cone. For an observer in $M$-brane universe, a causal effect can be  re-interpreted as a space-like event on ${\bar M}$-brane universe. The de-linked effective actions are given by 
\rs
\bea
S_M&\rightarrow&\int d^{11}x\ {\sqrt{-G}}\ \Big [\Big ( R_G\ -{1\over{48}}{\cal F}^a_4N_{ab}{\cal F}^b_4\Big )\nonumber\\
&&\qquad +\ \Big ({\cal H}_3\wedge{\cal F}_4\wedge {\cal F}_4\ +\ {\cal B}_3\wedge F_4\wedge F_4\ \Big )\Big ]\nonumber\\
{\rm and}\;\ S_{\bar M}&\rightarrow&\int d^{11}{\bar x}\ {\sqrt{-{\bar G}}}\ \Big ( {\cal K}\ -{1\over{240}}F_5^2\Big )\ .\label{form-5}
\eea

\rs
\noindent
In addition to an emergent $M$-theory, a de-linked $M'$-theory in $d$$=$$11$ representing the ${\bar M}$-brane is remarkable. It is important to observe that the $M'$-theory is not a NP-theory as it does not govern a propagating ${\cal H}_3$. 
Further investigation may reveal a plausible map between a $M'$-theory on $S^1$ and the $d$$=$$10$ type IIB (super)string theory.
The low energy limit is identified as a decoupling limit as the NP-dynamical correction decouples there. The ${\cal H}_3$-dynamics freezes in the limit, $i.e.\ {\cal H}_3\rightarrow {\cal H}_3^0$ (a constant). Then, the $M$-brane action becomes
\rs
\be
S_M=\int d^{11}x\Big [{\sqrt{-G}}\ \Big (R_G -{1\over{48}}F_4^2\Big )+ B_3\wedge F_4\wedge F_4\Big ]\ .\label{form-6}
\ee

\rs
\noindent
The decoupled emergent $M$-brane may formally be identified with the bosonic sector of $d$$=$$11$ SUGRA \cite{cremmer-julia-scherk} in natural units. The presence of fermionic local degrees are not affected by the decoupling of a boson dynamics. 
In fact, the hidden scalar $\phi$ dynamics (\ref{form-4}) is believed to break the local supersymmetry in $d$$=$$11$. 
Arguably the absence of $\phi$-dynamics in a limit restores the supersymmetry and may lead to the SUGRA theory. The low energy limit of $M$-brane is consistent with the $M$-theory in the same limit where it is known to describe the $d$$=$$11$ SUGRA. The result provokes thought to believe that a geometric torsion is a plausible candidate to describe a complete NP-theory of quantum gravity. Presumably it gets an interpretation for an {\it unified-quantum}. It provokes thought to speculate that a NP-theory can be a potential candidate for an unified gauge theory \cite{wilczek} of all four fundamental interactions. It may resolve some of the mystries related to the origin of dark energy in the  universe. A result that the GTR may be realized in a decoupling dynamics of a geometric torsion in superstring theory is new and is believed to enlighten the legacy of non-perturbation gravity to its depth.

\sp
\sp
\noindent
{\it\bf Acknowledgements:}
Author gratefully thank Josheph Polchinski, Fernando Quevedo and John H. Schwarz for insightful discussions at an early stage of the research. The work is partly supported by the University of Delhi R$\&$D grant RC/2015-16/9677.

\def\anp{Ann. Phys.}
\def\cmp{Comm. Math. Phys.}
\def\springer{Springer. Proc. Phys.}
\def\prl{Phys. Rev. Lett.}
\def\prd#1{{Phys. Rev.} {\bf D#1}}
\def\jhep{JHEP\ {}}{}
\def\cqg{Class. \& Quant. Grav.}
\def\plb#1{{Phys. Lett.} {\bf B#1}}
\def\npb#1{{Nucl. Phys.} {\bf B#1}}
\def\mpl#1{{Mod. Phys. Lett} {\bf A#1}}
\def\ijmpa#1{{Int. J. Mod. Phys.} {\bf A#1}}
\def\ijmpd#1{{Int. J. Mod. Phys.} {\bf D#1}}
\def\mpla#1{{Mod. Phys. Lett.} {\bf A#1}}
\def\rmp#1{{Rev. Mod. Phys.} {\bf 68#1}}
\def\jaat{J.Astrophys.Aerosp.Technol.\ {}} {}
\def \epj#1{{Eur. Phys. J.} {\bf C#1}} 
\def \jcap{JCAP\ {}}{}

\end{document}